\documentclass[3p,twocolumn,number,sort&compress]{elsarticle}
\usepackage{lineno}
\usepackage{natbib}
\usepackage{hyperref}
\usepackage{graphics}
\usepackage{amsfonts}
\usepackage{amssymb,amsmath}

\begin{document}

\begin{frontmatter} 

\title{Development of Combined Opto-Acoustical Sensor Modules}

\author{A.~Enzenh\"ofer\corref{cor1}}
\ead{alexander.enzenhoefer@physik.uni-erlangen.de}
\ead[url]{http://www.acoustics.physik.uni-erlangen.de}

\author{G.~Anton}
\author{K.~Graf}
\author{J.~H\"o\ss l}
\author{U.~Katz}
\author{R.~Lahmann}
\author{M.~Neff}
\author{C.~Richardt}

\cortext[cor1]{Corresponding author}
\address{Friedrich-Alexander-Universit\"at Erlangen-N\"urnberg, Erlangen Centre for Astroparticle Physics, Erwin-Rommel-Str.\,1, D-91058 Erlangen, Germany}

\begin{abstract}

The faint fluxes of cosmic neutrinos expected at very high energies require large instrumented detector volumes.
The necessary volumes in combination with a sufficient shielding against background constitute forbidding and complex environments (e.g.~the deep sea) as sites for neutrino telescopes.
To withstand these environments and to assure the data quality, the sensors have to be reliable and their operation has to be as simple as possible.
A compact sensor module design including all necessary components for data acquisition and module calibration would simplify the detector mechanics and ensures the long term operability of the detector.
The compact design discussed here combines optical and acoustical sensors inside one module, therefore reducing electronics and additional external instruments for calibration purposes.
In this design the acoustical sensor is primary used for acoustic positioning of the module.
The module may also be used for acoustic particle detection and marine science if an appropriate acoustical sensor is chosen.\\
First tests of this design are promising concerning the task of calibration.
To expand the field of application also towards acoustic particle detection further improvements concerning electromagnetic shielding and adaptation of the single components are necessary.

\end{abstract}


\end{frontmatter}

\linenumbers

\section{Introduction}
\label{sec:introduction}

Cosmic neutrinos as messenger particles of processes at the highest energies are very important for our understanding of the underlying physical processes.
They are the only viable messengers for extragalactic sources as the range of other messenger particles (photons, electrons, protons and nuclei) is confined by interactions with the intergalactic medium.
The small cross section of the neutrinos in combination with the faint fluxes expected for high energy particles lead to the requirement of large detector volumes and corresponding huge amounts of sensors.
These necessary detector sizes are not affordable with common sensor designs, so the improvement of existing as well as development of new sensors and detection techniques is mandatory.
Existing or planned research infrastructures are test sites for these new developments.
In this context new approaches like acoustic detection and radio detection techniques are pursued.
Besides detectors based on only one detection technique, efforts are ongoing to combine different techniques.

Especially dynamic environments like the deep sea constitute forbidding and complex environments for detector operation.
The detector elements cannot be fixed in space so the sensor position and orientation requires permanent monitoring.
This task is performed e.g.~in the neutrino Cherenkov telescope ANTARES \cite{ant_col} with an acoustic positioning system \cite{ant_pos} consisting of a dedicated set of acoustic emitters and receivers.
These acoustic emitters and receivers are distributed all over the detector to achieve sufficient coverage.
In addition, compasses and tiltmeters complement the information of the acoustic positioning system.

This set-up was taken as starting point for considerations about combining optical and acoustical sensors in one module.
The experiences gained in the operation of ANTARES and AMADEUS \cite{amadeus_nimA} and other deep-sea infrastructures also revealed that cable feedthrough are a vulnerable point of the design.
Reducing the number of feedthrough will reduce the probability for failures of the sensors due to water ingress.
This and a further reduction of costs by confining the amount of used electronics and the use of multi-purpose devices can be achieved by combined sensor modules.
Basic requirements of such combined sensor modules may include:

\begin{itemize}
\item
Compact design.
\item
Single power supply.
\item
One data acquisition chain for all sensors.
\item
Self-contained calibration for the module.
\item
Variably utilisable data, e.g.~also for multidisciplinary purposes.
\end{itemize}

For future deep-sea neutrino telescopes, for example KM3NeT \cite{KM3NET}, these combined sensor modules could provide unique properties, e.g.~ability for inherent position and orientation calibration, an enhanced possibility to study the deep-sea environment (encourages multidisciplinarity) and the use of complementary neutrino detection methods.
In large-scale projects the aspect of multidisciplinarity is very important as the available resources are restricted.

This article focuses on the first steps towards the realisation of a combined module performed at the Erlangen Centre for Astroparticle Physics (ECAP) \cite{ECAP}.

\section{Module concept}
\label{sec:concept}

ECAP is partner in the ANTARES Collaboration and the KM3NeT Consortium.
The ANTARES neutrino telescope can be seen as a predecessor for the future large-scale neutrino telescope KM3NeT.
Within the ANTARES infrastructure it is possible to test detector hardware and software in order to prove and improve its performance.
In this context ECAP follows a leading role in the acoustic detection test set-up AMADEUS as well as in the acoustic positioning system.
The combination of expertise in optical and acoustic particle detection foster considerations of a possible combination of both techniques inside one detection module.
The basic concept of a so-called {\bf O}pto-{\bf A}coustical {\bf M}odule (OAM) is the combination of acoustical and optical sensors in one housing.
This combination provides advantages for the resulting module and additional options:

\begin{itemize}
\item
Reduction of costs because of less mechanics, sensors and electronics.
\item
Reduction of cable feedthrough, which are possible weak points in deep-sea operation.
\item
Shared use of electronics which simplifies the construction and data acquisition of the module.
\item
The acoustical sensor can also be used for calibration purposes.
\item
Extend the research activities also to marine sciences and enhance multidisciplinary cooperation e.g.~with marine scientists.
\item
Combination of complementary detection techniques to extend the accessible energy range.
\end{itemize}

The main difficulties originate in the limited space inside the module and the interference between the sensors.
Primarily the high voltage generation and the electronics-noise-rich PMT environment renders the combined operation problematic.

\section{Prototype and test set-up}
\label{sec:prototype_test_set-up}

\subsection{Prototype}
\label{subsec:prototype}

A prototype OAM was built at ECAP in order to test the possible operation of both detection techniques inside the same sensor module.
This prototype comprises the following components:

\begin{itemize}
\item
A 10" PMT with an active base to generate the necessary high voltages from a single power supply.
Both are identical to the ones used in ANTARES \cite{ant_pmt}.
\item
An acoustical sensor consisting of a piezo ceramic with custom designed preamplifier as used in the Acoustic Modules deployed in one out of six AMADEUS Acoustic Storeys for feasibility studies \cite{amadeus_nimA}.
\item
The lower 17" glass hemisphere of a standard ANTARES Optical Module \cite{oms}.
\end{itemize}

The piezo ceramic is glued to the inside of the hemisphere next to the PMT.
The PMT is fixed with two plastic discs replacing the optical gel and other mechanical parts normally used to fix the PMT inside the module.
No particular electromagnetic shielding is applied to both sensors.
This simple design was chosen in order to preserve high flexibility and to simplify the test of different configurations.
In addition the standard way of integrating the sensors is time consuming and difficult and complicates the adoption of improvements.

\subsection{Test set-up}
\label{subsec:test_set-up}
The OAM prototype is put inside a grounded metal box to reduce the electromagnetic noise originating from the laboratory environment.
The metal box is additionally blackened inside and taped light-proof to enable a PMT operation without optical background.
To reduce acoustic coupling between module and box, the module is placed on rubber foam.
A LED is placed next to the OAM to provide a signal to the PMT.
This trigger is a simple emulation of nominal PMT working conditions.
Each component of the prototype can be switched on or off and operated from outside the box.
In addition some adjustments of parameters can be performed in specified ranges.

\section{First results}
\label{sec:results}

\begin{figure}
\includegraphics[width=\linewidth]{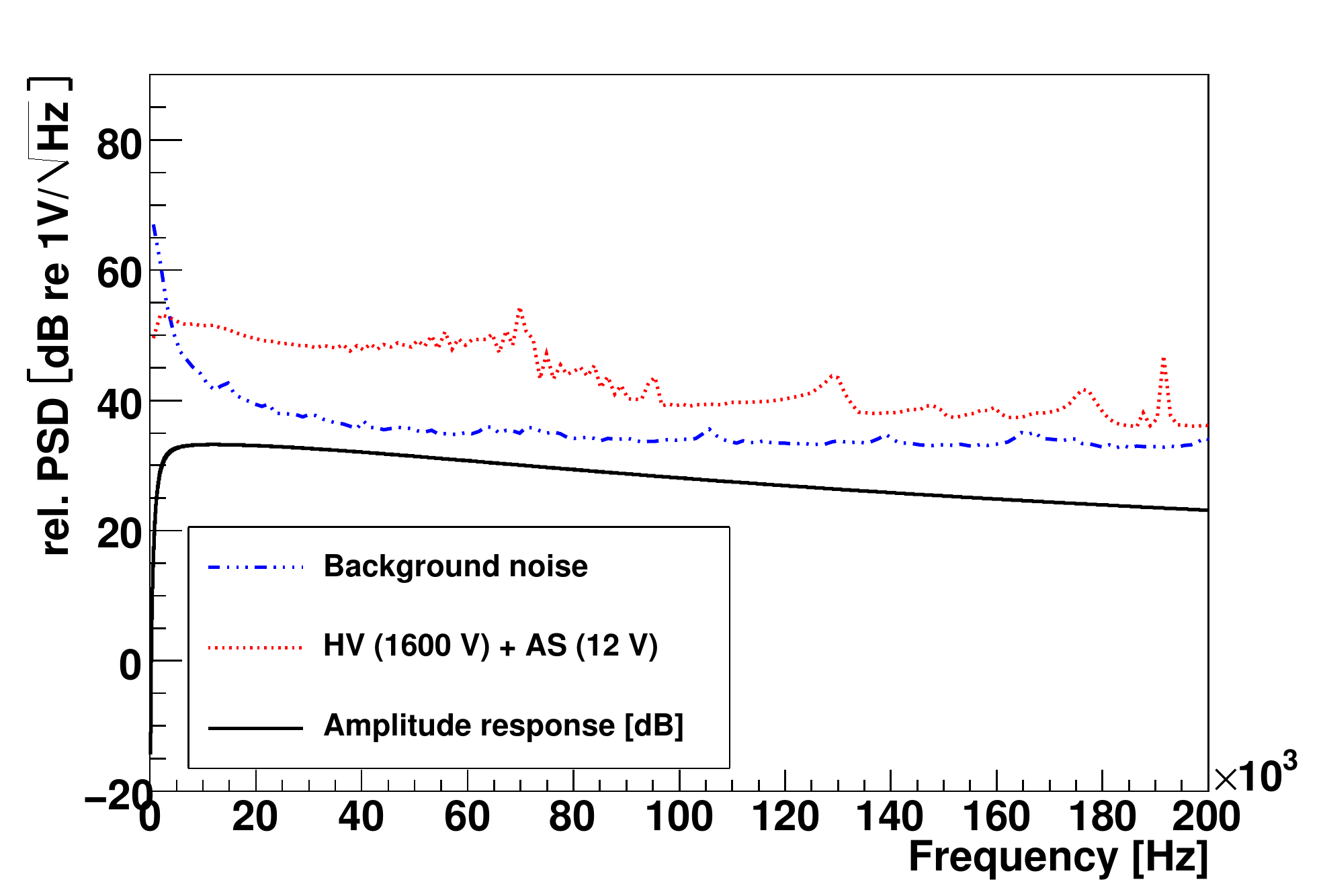}
\caption[Background noise, amplitude response of the preamplifier and first measurement]{Power spectral density (PSD) for the conditions in the laboratory (dash-dotted line) together with the amplitude response (solid line) of the acoustic preamplifier. The dotted line shows the result for the setting given in the legend (AS stands for the preamplifier of the {\bf a}coustical {\bf s}ensor). Further description is given in the text.}
\label{fig:result_1}
\end{figure}

\begin{figure}
\includegraphics[width=\linewidth]{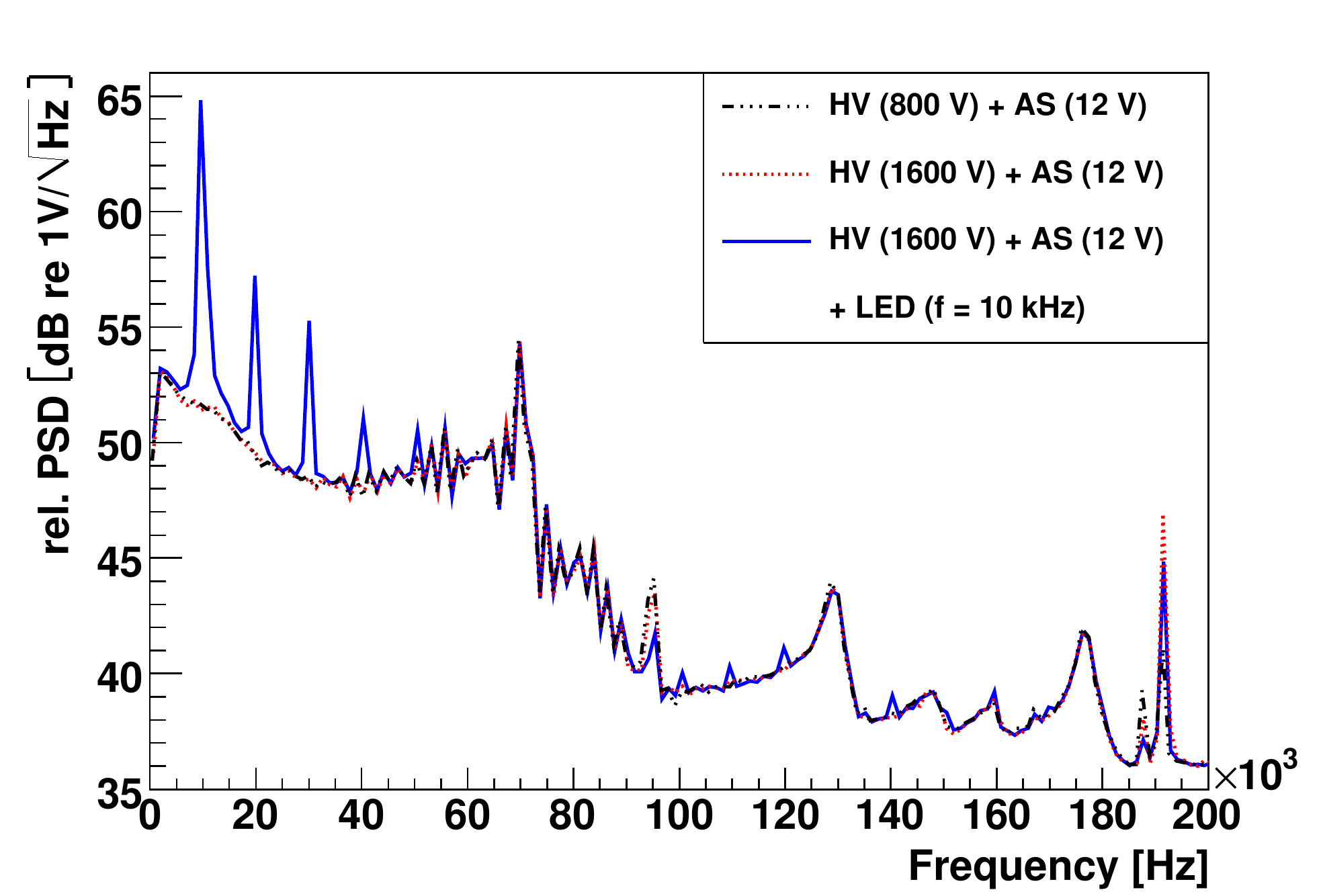}
\caption[Test results for different test settings]{PSD for different test settings. These settings include the minimum settable high voltage (dash-dotted line), medium high voltage (dotted line) and the latter setting with a flashing LED as given in the legend (AS stands for the preamplifier of the {\bf a}coustical {\bf s}ensor). The dotted line is identical to the one in Figure \ref{fig:result_1} for reference purposes. Further description is given in the text.}
\label{fig:result_2}
\end{figure}

Figure \ref{fig:result_1} and Figure \ref{fig:result_2} show the first results gained in the performed tests with the help of a digital oscilloscope.
The dash-dotted line in Figure \ref{fig:result_1} corresponds to the background noise in the laboratory.
This background was recorded over the acoustical sensor inside the metal box with the fully connected but unpowered module.
It shows an almost flat spectrum for higher frequencies but it raises towards decreasing frequencies below 40\,kHz.
The dotted line shows the measured noise spectrum recorded by the powered acoustical sensor with powered but untriggered PMT.
The preamplifier of the acoustical sensor is powered by 12\,V, which is the nominal voltage for the preamplifier used in this measurement.
The PMT in this case is operated with 1600\,V between cathode and anode.
This voltage lies well within the PMT power supply range from 1300\,V to 2300\,V.
Figure \ref{fig:result_1} also provides the amplitude response of the preamplifier used (solid line) for explanatory reason.
Its behaviour is clearly visible throughout the whole frequency range:
The band-pass characteristic removing the low frequency part of the noise as well as a global decrease towards higher frequencies.
Besides this global behaviour of the noise, explainable through the preamplifier characteristics, some additional, localised pattern occur due to:

\begin{itemize}
\item
The piezo ceramic itself and its coupling to the glass sphere.
This influence is responsible for the spiked structure between 40\,kHz and 90\,kHz.
\item
The influence of the high voltage generation inside the PMT base.
This influence manifests itself in some prominent peaks e.g.~at about 92\,kHz, 130\,kHz, 180\,kHz or 190\,kHz.
\end{itemize}

Figure \ref{fig:result_2} depicts the results for two different test settings:
The dash-dotted line is recorded with the minimum voltage settable for the PMT, i.e.~800\,V between cathode and the first dynode and no voltage between the other electrodes, whereas the solid line is recorded with 1600\,V between cathode and anode.
In the latter case the PMT is triggered by a LED flashing with a frequency of 10\,kHz.
The dotted line is shown for comparison and is identical to the dotted line in Figure \ref{fig:result_1}.
The reduction of the high voltage has no significant influence on the noise level except some changes at the specific frequencies already mentioned above.
The influence of the flashing LED is clearly visible through high peaks at multiple frequencies of 10\,kHz.
The peaks are very dominant but their integrated power content is small compared to the whole frequency spectrum.
This result is no real drawback for positioning purposes as the necessary signals well exceed this level.
The prototype is not applicable for acoustic particle detection in the current design as the expected acoustic signals are to weak.
It has to be pointed out that the acoustical sensor used for this tests was not optimised for the use in combination with a PMT.
Mainly the preamplifier allows for some optimisation which is currently in progress.
The test is also a benchmark of a worst case set-up without proper shielding against electromagnetic influences inside the module.
This shielding will be optimised as soon as the preamplifier design has been finished.

\section{Conclusions}
\label{sec:conclusions}

The general operability of both optical and acoustical sensor techniques inside the same module is possible even in a very simple design as presented in this article.
The Opto-Acoustical Module (OAM) can be used for optical detection, calibration purposes and to study the deep-sea environment.
To further improve the results as well as to enlarge the area of application of the OAM also towards acoustic neutrino detection, further improvements are pursued.
These improvements are mainly focused on electromagnetic shielding which is necessary to reduce the inter-sensor crosstalk.
This interference also has to be evaluated and studied carefully for the influence of the acoustical sensor operation on the PMT operation.
This influence should be negligible due to the relatively low voltages and frequencies of the acoustical sensor involved, but should be evaluated prior to a final design.
Further improvements on the signal preamplification and further signal processing are mandatory to exploit the full potential of these modules.
Tests are planned with improved designs as well as complete data acquisition chains particularly with regard to the future km$^{3}$-scale neutrino telescope KM3NeT.

\section*{Acknowledgements}
\label{sec:acknowledgements}

The authors acknowledge the support by the German government through BMBF grants 05CN5WE1/7 and 05A08WE1.


{}

\end{document}